\documentclass[prb,twocolumn,showpacs,amsmath,amssymb,superscriptaddress]{revtex4-2}

\usepackage{amsmath,amssymb,amsfonts,bm,color,graphicx,tabularx}

\usepackage[unicode=true,colorlinks=true]{hyperref}

\usepackage[etex=true,export]{adjustbox}
\usepackage{makecell}
\usepackage{multirow}

\hypersetup{linkcolor=blue,citecolor=blue,urlcolor=blue}

\usepackage{tabularx,multirow,array,diagbox}
\usepackage{adjustbox}

\newcommand{\beq}{\begin{equation}}
\newcommand{\eeq}{\end{equation}}
\newcommand{\beql}{\begin{equation*}}
\newcommand{\eeql}{\end{equation*}}
\newcommand{\beqn}{\begin{eqnarray}}
\newcommand{\eeqn}{\end{eqnarray}}

\begin{document}
\title{Family of third-order topological insulators from Su-Schrieffer-Heeger stacking}
\author{Xun-Jiang Luo}
\affiliation{School of Physics and Technology, Wuhan University, Wuhan 430072, China}
\affiliation{Department of Physics, Hong Kong University of Science and Technology, Clear Water Bay, Hong Kong, China}
\author{Jia-Zheng Li}
\affiliation{School of Physics and Technology, Wuhan University, Wuhan 430072, China}
\author{Meng Xiao}
\affiliation{School of Physics and Technology, Wuhan University, Wuhan 430072, China}
\affiliation{Wuhan Institute of Quantum Technology, Wuhan 430206, China}
\author{Fengcheng Wu}
\email{wufcheng@whu.edu.cn}
\affiliation{School of Physics and Technology, Wuhan University, Wuhan 430072, China}
\affiliation{Wuhan Institute of Quantum Technology, Wuhan 430206, China}

\begin{abstract}
We construct a family of chiral symmetry-protected third-order topological insulators by stacking Su-Schrieffer-Heeger (SSH) chains and provide a unified topological characterization by a series of Bott indices. Our approach is informed by the analytical solution of corner states for the model Hamiltonians written as a summation of the extended SSH model along three orthogonal directions. By utilizing the generalized Pauli matrices, an enumeration of the constructed model Hamiltonians generates ten distinct models, including the well-studied three-dimensional Benalcazar-Bernevig-Hughes model. By performing a boundary projection analysis for the ten models, we find that certain surfaces and hinges of the systems can exhibit, respectively, nontrivial second-order and first-order topology in the phase of the third-order topological insulators. Furthermore, we analyze the phase diagram for one of the predicted models and reveal a rich set of topological phases, including the third-order topological insulators, second-order weak topological insulators, and second-order nodal semimetals.

\end{abstract}


\maketitle

\begin{figure*}
\centering
\includegraphics[width=7in]{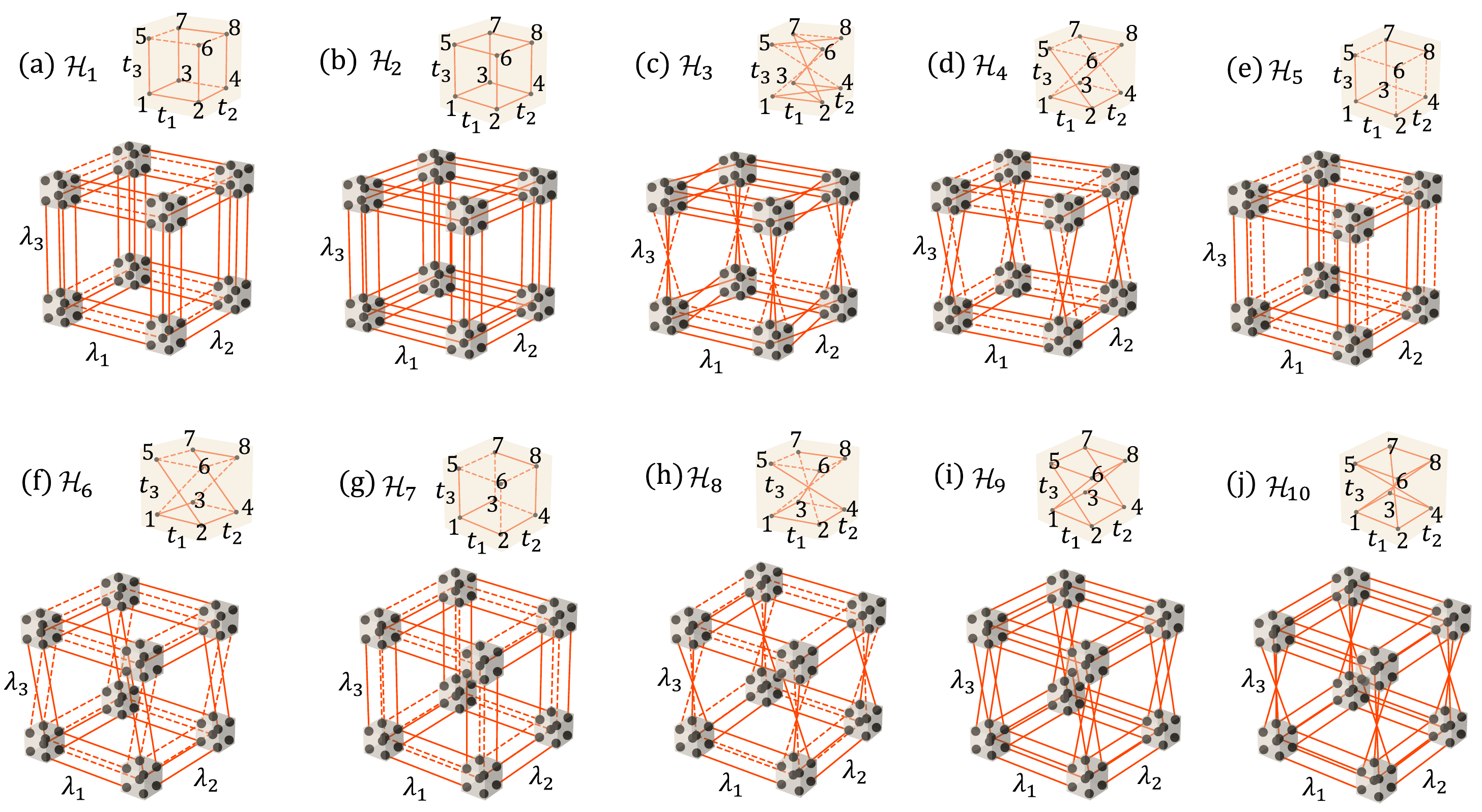}
\caption{(a)-(j) Intercellular hoppings for the Bloch
Hamiltonians $\mathcal{H}_{1,\cdots,10}$ in Table~\ref{tab2}.  The insets in (a)-(j) show the degrees of freedom and hoppings within the
unit cell.  The dashed lines represent negative hopping parameters.}
\label{Fig1}
\end{figure*}

\section{Introduction}

Higher-order topological phases \cite{Benalcazar2017a,Benalcazar2017,Song2017,Langbehn2017}, encompassing  insulators \cite{Schindler2018,Khalaf2018,Geier2018,Ezawa2018,Schindler2018a,Trifunovic2019,Wang2019,Xu2019,Park2019,Sheng2019,Zhang2020b,2021khalaf}, superconductors \cite{Yan2018,Zhu2018,Wang2018,Wang2018a,Zhang2019,Zhu2019,Zhang2019a,Yan2019,Pan2019,Wu2020,Luo2021a,Pan2022,Luo2024}, and semimetals \cite{TopologicalQuadrupolarSemimetals2018lina,HigherOrderWeylSemimetals2020ghorashi,StrongFragileTopological2020wieder,Wang2020,UniversalHigherorderBulkboundary2022lenggenhager}, have garnered substantial research interest. In general, an 
$n$th-order topological phase in a 
$d$ dimensional ($d$D) system is characterized by the presence of topologically protected gapless boundary states at its ($d-n$)D boundaries. For instance, a 
$d$th-order topological insulator in a 
$d$D system exhibits robust zero-energy corner states (ZECSs) that are protected by both bulk and boundary energy gaps. These ZECSs can lead to the emergence of fractional corner charges in insulators \cite{Benalcazar2019} and non-Abelian anyons in superconductors \cite{Zhang2020c}, highlighting their potential applications in topological quantum computation \cite{Nayak2008}.

A prominent example of higher-order topological insulators is the 2D BBH model \cite{Benalcazar2017a,Benalcazar2017}, which features a bulk quadrupole moment and has been experimentally realized in various artificial lattice systems, including phononic crystals \cite{Serra-Garcia2018,Mittal2019}, acoustic crystals \cite{Qi2020}, and electrical circuits \cite{Serra-Garcia2019,Imhof2018}. The BBH model can be interpreted as a stacking of SSH \cite{Su1979} chains, thereby serving as a higher-dimensional generalization of the 1D SSH model.
Particularly, the SSH chains can be coupled through different hopping configurations along the stacking direction, resulting in a diverse range of models \cite{Luo2022,Lichangan2022}. For instance, in addition to the BBH model, the 2D SSH model \cite{Liu2017} and the 2D crossed SSH model \cite{Luo2022}, both exhibiting second-order topology, can also be realized through specific configurations of SSH stacking.  The model Hamiltonians of these three models can be written in a unified manner as a summation of the extended SSH model, which is multiple copies of the SSH model, along two directions.
This formulation suggests that we can investigate SSH stacking systems starting from the momentum space Bloch Hamiltonians.

The quest for a comprehensive topological characterization of higher-order topological phases has been a focal point of research in recent years \cite{Benalcazar2017a,Benalcazar2017,Liu2017,Khalaf2018a,Kang2019,Wheeler2019,Benalcazar2019,20LiHe,Li2020a,Ren2020,20Araki,Benalcazar2022,22Wienand,Luo2022,23Tada,23Dubinkin,24Jahin,24Lin}. Various higher-order topological invariants have been proposed to characterize corner states, including the nested Wilson loop \cite{Benalcazar2017,Benalcazar2017a}, the multipole moment \cite{Wheeler2019,Benalcazar2019}, and the chiral multipole number \cite{Benalcazar2022}. However, these invariants do not offer a universal characterization for corner states, which can be examined in the SSH stacking 2D systems \cite{Luo2022}. Consequently, the SSH stacking systems provide not only a platform for realizing higher-order topological phases but also a testbed to evaluate the applicability of the proposed topological invariants.

In this work, we construct a family of chiral symmetry-protected third-order topological insulators (TOTIs), described by model Hamiltonians expressed as a summation of the extended SSH model along three orthogonal directions. Utilizing an analytical approach to construct corner states and an enumeration of the
model Hamiltonians,  we identify ten distinct models with lattice hoppings illustrated in Fig.~\ref{Fig1}. These models realize TOTIs in the appropriate parameter regimes where the bulk, surface, and hinge spectra are all gapped. In the phase of the TOTIs, through a boundary projection analysis of these ten models, we extract the effective boundary Hamiltonians for certain surfaces (hinges), which are found to be equivalent to the 2D BBH, 2D SSH, or 2D crossed SSH models \cite{Luo2022} (the two-band SSH model). Our boundary projection analyses elucidate the hierarchy inherent in these systems. Building upon the theoretical framework established in Ref. \onlinecite{Jiazheng2024}, we characterize the corner states in a unified way by using a series of Bott indices, which effectively capture the chirality configurations of corner states in real space. Finally, we present a rich phase diagram for one of the predicted models, which includes TOTIs, second-order weak topological insulators, and second-order nodal semimetals.

The rest of this paper is organized as follows. In Sec.~\ref{II}, we 
construct a class of model Hamiltonians and construct their ZECSs solution analytically. In Sec.~\ref{III}, we enumerate the constructed 3D model Hamiltonians by utilizing generalized Pauli matrices, which gives rise to ten distinct models that can exhibit third-order topology.
In Sec.~\ref{IV}, we perform boundary projection analysis for the ten models and extract their boundary Hamiltonians. In Sec.~\ref{V}, we characterize the higher-order topology by a series of Bott indices. In Sec.~\ref{VI}, we analyze the phase diagram for one of the predicted Hamiltonians. In Sec.~\ref{VII}, we conclude with a discussion and summary. Appendices \ref{appendixA}-\ref{appendixD} complement the main text with additional technical details.

\section{Stacking of SSH model and Corner States}
\label{II}

To describe the SSH stacking systems, we  construct the $d$D Bloch Hamiltonian
\beqn
\begin{aligned}
&H_d(\bm k_d)=\sum_{s=1}^dh_s(k_s),\\
&h_s(k_s)=M_s(k_s)\Gamma_{sa}^{(d)}+\lambda_s\sin k_s\Gamma_{sb}^{(d)},
\label{h1}
\end{aligned}
\eeqn
where the momentum vector is $\bm k_d=(k_1,\cdots,k_d)$. $M_s(k_s)$ is $t_s+\lambda_s\cos k_s$ with $t_s$ and $\lambda_s$ being the model parameters. Matrices $\Gamma_{sa}^{(d)}$ and $\Gamma_{sb}^{(d)}$ are the generalized Pauli matrices represented by a direct product of $d$ copies of Pauli matrices ( see Appendix \ref{appendixA}) and satisfy the anti-commuting relation $\{\Gamma_{sa}^{(d)},\Gamma_{sb}^{(d)}\}=0$, with $s=1,\cdots,d$.
Therefore, the 1D Hamiltonian $h_s$ respects the chiral symmetry $C_s=-i\Gamma_{sa}^{(d)}\Gamma_{sb}^{(d)}$ with $C_s^2=1$. When $d=1$, $h_s$ is exactly the two-band SSH model.
When $d>1$, $h_s$ can be considered as the direct sum of $2^{d-1}$ copies of the two-band SSH model. Thus, we dub $h_s$ the extended SSH model \cite{Luo2022,Luo2023} and $H_d(\bm k_d)$ can be viewed as a higher-dimensional generalization of the SSH model.

When $|t_s|<|\lambda_s|$, $h_s$ is topologically nontrivial and hosts $2^{d-1}$ zero-energy states localized at each end under the open boundary condition.
The wave function of these end zero-energy states can be obtained by solving the eigen equation $h_s(r_s)|\Phi_{\alpha_s}^{(s)}(r_s)\rangle=0$ (see Appendix \ref{appendixB}), with $h_s(r_s)$ being the real space Hamiltonian of $h_s$. $|\Phi_{\alpha_s}^{(s)}(r_s)\rangle$ is the eigenstate of $C_s$ with eigenvalue $z_s=\pm 1$. The end states labeled by $z_s=-1$ and $z_s=1$ are localized close to $r_s=-L_s/2$ and $r_s=L_s/2$ respectively, with $L_s$ being the length of the system along the $r_s$ direction. Thus, the 1D end zero-energy states wave function of $h_s$ can be written as
\beqn
|\Phi_{z_s}^{(s)}(r_s)\rangle=\mathcal{N}f_{z_s}^{(s)}(r_s)|\psi_{z_s}^{(s)}\rangle,
\label{es}
\eeqn
where the scalar function $f_{-(+)}^{(s)}(r_s)$ is localized close to the left (right) end, the spinor $|\psi_{z_s}^{(s)}\rangle$ satisfies $C_s|\psi_{z_s}^{(s)}\rangle=z_s|\psi_{z_s}^{(s)}\rangle$, and $\mathcal{N}$ is the normalization factor.

\begin{table}[htb]
\centering
\setlength\tabcolsep{1.5pt}
\renewcommand{\multirowsetup}{\centering}
\renewcommand{\arraystretch}{1.1}
\caption{ Ten types of inequivalent relations between  matrices $\{\Gamma_{1a,2a,3a}^{(3)},\Gamma_{1b,2b,3b}^{(3)}\}$. In case (1), $\text{w}=\text{w}^{\prime}=\text{w}^{\prime\prime}$ for $(12\text{w},13\text{w}^{\prime},23\text{w}^{\prime\prime})$. In case (2), $\text{w}\neq \text{w}^{\prime}\neq \text{w}^{\prime\prime}$. In case (3), $\text{w}= \text{w}^{\prime}\neq \text{w}^{\prime\prime}$, or $\text{w}=\text{w}^{\prime\prime}\neq \text{w}^{\prime}$, or $\text{w}^{\prime}= \text{w}^{\prime\prime}\neq \text{w}$. }
\begin{tabular}{|c|c|}
\hline
(1) & (12i,23i,13i), (12ii,23ii,13ii),(12iii,23iii,13iii), \\
\hline
(2)&(12i,23ii,13iii),\\
\hline
(3)&\makecell{(12i,23i,13ii),(12i,23i,13iii),(12i,23ii,13ii),(12i,23iii,13iii),\\
(12ii,23ii,13iii),(12ii,23iii,13iii),}\\
\hline
\end{tabular}
\label{tab1}
\end{table}

\begin{table*}[htb]
\centering
\setlength\tabcolsep{2.3pt}
\renewcommand{\arraystretch}{1.3}
\caption{Matrix representation of  the ten types of relations $(12\text{w},13\text{w}^{\prime},23\text{w}^{\prime\prime})$ presented in Table~\ref{tab1}. Matrix $\Sigma_{ijk}$ is $\Sigma_{ijk}=\sigma_i\otimes\sigma_j\otimes\sigma_k$ with $i,j,k=0,1,2,3$.
Operator $\mathcal{C}$ denotes the chiral symmetry of Hamiltonian $\mathcal{H}_i$, which anti-commutes with the matrices  $\{\Gamma_{1a,2a,3a}^{(3))},\Gamma_{1b,2b,3b}^{(3))}\}$. The Operator $C_{s}$ is defined by $C_s=i\Gamma_{sa}^{(3)}\Gamma_{sb}^{(3)}$ with $s=1,2,3$.}
\begin{tabular}{|c|c|c|c|c|c|c|c|c|c|c|c|c|c|}
\hline
Third-order topological phases&Models&   types& $\Gamma_{1a}^{(3)}$ & $\Gamma_{1b}^{(3)}$ & $\Gamma_{2a}^{(3)}$ &$\Gamma_{2b}^{(3)}$  &$\Gamma_{3a}^{(3)}$ &$\Gamma_{3b}^{(3)}$ &$C_1$&$C_2$&$C_3$&$\mathcal{C}$&relation\\
\hline
\multirow{9}{*}{\makecell{$\mathcal{H}_i=\sum_{s=1}^3h_s(k_s)$,\\$h_s=M_s(k_s)\Gamma_{sa}^{(3)}+\lambda_s\sin k_s\Gamma_{sb}^{(3)}$}}&$ \mathcal{H}_1$&(12i,23i,13i)& $\Sigma_{331}$&$\Sigma_{332}$ &$\Sigma_{310}$ &$\Sigma_{320}$&$\Sigma_{100}$&$\Sigma_{200}$&$\Sigma_{003}$&$\Sigma_{030}$&$\Sigma_{300}$&$\Sigma_{333}$&$\mathcal{C}=C_1C_2C_3$\\
\cline{2-14}
~&$ \mathcal{H}_2$&(12ii,23ii,13ii)& $\Sigma_{001}$&$\Sigma_{002}$ &$\Sigma_{010}$ &$\Sigma_{020}$&$\Sigma_{100}$&$\Sigma_{200}$&$\Sigma_{003}$&$\Sigma_{030}$&$\Sigma_{300}$&$\Sigma_{333}$&$\mathcal{C}=C_1C_2C_3$\\
\cline{2-14}
~& $ \mathcal{H}_3$& (12iii,23iii,13iii)& $\Sigma_{310}$&$\Sigma_{320}$ &$\Sigma_{011}$ &$\Sigma_{012}$&$\Sigma_{122}$&$\Sigma_{222}$&$\Sigma_{030}$&$\Sigma_{003}$&$\Sigma_{300}$&$\Sigma_{030}$&$\mathcal{C}=C_1$\\
\cline{2-14}
~&$\mathcal{H}_4$&(12i,13ii,23iii)& $\Sigma_{331}$&$\Sigma_{332}$ &$\Sigma_{310}$ &$\Sigma_{320}$&$\Sigma_{132}$&$\Sigma_{232}$&$\Sigma_{003}$&$\Sigma_{030}$&$\Sigma_{300}$&$\Sigma_{033}$&$\mathcal{C}=C_1C_2$\\
\cline{2-14}
~&$ \mathcal{H}_5$&(12i,23i,13ii)& $\Sigma_{331}$&$\Sigma_{332}$ &$\Sigma_{310}$ &$\Sigma_{320}$&$\Sigma_{103}$&$\Sigma_{203}$&$\Sigma_{003}$&$\Sigma_{030}$&$\Sigma_{300}$&$\Sigma_{333}$&$\mathcal{C}=C_1C_2C_3$\\
\cline{2-14}
~&$ \mathcal{H}_6$&(12i,23i,13iii)& $\Sigma_{331}$&$\Sigma_{332}$ &$\Sigma_{310}$ &$\Sigma_{320}$&$\Sigma_{102}$&$\Sigma_{202}$&$\Sigma_{003}$&$\Sigma_{030}$&$\Sigma_{300}$&$\Sigma_{033}$&$\mathcal{C}=C_1C_2$\\
\cline{2-14}
~&$\mathcal{H}_7$&(12i,23ii,13ii)& $\Sigma_{331}$&$\Sigma_{332}$ &$\Sigma_{310}$ &$\Sigma_{320}$&$\Sigma_{133}$&$\Sigma_{233}$&$\Sigma_{003}$&$\Sigma_{030}$&$\Sigma_{300}$&$\Sigma_{333}$&$\mathcal{C}=C_1C_2C_3$\\
\cline{2-14}
~&$ \mathcal{H}_8$&(12i,23iii,13iii)& $\Sigma_{331}$&$\Sigma_{332}$ &$\Sigma_{310}$ &$\Sigma_{320}$&$\Sigma_{121}$&$\Sigma_{221}$&$\Sigma_{003}$&$\Sigma_{030}$&$\Sigma_{300}$&$\Sigma_{333}$&$\mathcal{C}=C_1C_2C_3$\\
\cline{2-14}
~&$\mathcal{H}_9$&(12ii,13ii,23iii)& $\Sigma_{001}$&$\Sigma_{002}$ &$\Sigma_{010}$ &$\Sigma_{020}$&$\Sigma_{101}$&$\Sigma_{201}$&$\Sigma_{003}$&$\Sigma_{030}$&$\Sigma_{300}$&$\Sigma_{033}$&$\mathcal{C}=C_1C_2$\\
\cline{2-14}
~&$ \mathcal{H}_{10}$&(12ii,23iii,13iii)& $\Sigma_{001}$&$\Sigma_{002}$ &$\Sigma_{010}$ &$\Sigma_{020}$&$\Sigma_{111}$&$\Sigma_{211}$&$\Sigma_{003}$&$\Sigma_{030}$&$\Sigma_{300}$&$\Sigma_{333}$&$\mathcal{C}=C_1C_2C_3$\\
\hline
\end{tabular}
\label{tab2}
\end{table*}

Since $h_s$ hosts zero-energy end states, a natural question is whether $H_d$ has ZECSs under the open boundary conditions. Here, we present a sufficient condition for the existence of corner states. Namely, $\{C_1,\cdots, C_d\}$ commute with each other, expressed by $[C_s,C_{s^{\prime}}]=0$ for $s,s^{\prime}\in \{1,\cdots, d\}$. Under this condition, matrices $\{C_1,\cdots, C_d\}$ have $2^{d}$ common eigenstates $|\psi_{z_1,\cdots,z_d}\rangle$, where
\beqn
C_s|\psi_{z_1,\cdots,z_d}\rangle=z_s|\psi_{z_1,\cdots,z_d}\rangle,
\eeqn
with the eigenvalue $z_s=\pm 1$. Then we can construct $2^d$ corner states 
\beqn
|\Psi_{z_1,\cdots,z_d}(\bm r)\rangle=\mathcal{N}\prod_{s=1}^df_{z_s}^s(r_s)|\psi_{z_1,\cdots,z_d}\rangle.
\label{ZECS1}
\eeqn
It can be readily checked that 
\beqn
h_{s=1,\cdots,d}(r_s)|\Psi_{z_1,\cdots,z_d}(\bm r)\rangle=0,
\eeqn
which gives rise to $H_d(\bm r_d)|\Psi_{z_1,\cdots,z_d}(\bm r)\rangle=0$. Therefore, we can make a strong conclusion that  $H_d$ hosts $2^d$ ZECSs under the conditions
$[C_s,C_{s^{\prime}}]=0$ and $|t_s|<|\lambda_s|$ for $s,s^{\prime}\in \{1,\cdots, d\}$. In the following, we take $|t_s|<|\lambda_s|$ unless otherwise stated.

We note that the condition $[C_s,C_{s^{\prime}}]=0$ for $s,s^{\prime}\in \{1,\cdots, d\}$ can be satisfied in the two special cases described as follows. The first case is that all the matrices in Eq.~\eqref{h1} anti-commute with each other, for example,
\beqn
\begin{aligned}
&\Gamma_{sa}^{(d)}=\underbrace{\sigma_3\otimes\cdots\sigma_3}_{d-s}\otimes\sigma_2\otimes\underbrace{\sigma_0\cdots\sigma_0}_{s-1},\\
&\Gamma_{sb}^{(d)}=\underbrace{\sigma_3\otimes\cdots\sigma_3}_{d-s}\otimes\sigma_1\otimes\underbrace{\sigma_0\cdots\sigma_0}_{s-1},\\
&C_{s}=i\Gamma_{sa}^{(d)}\Gamma_{sb}^{(d)}=\underbrace{\sigma_0\otimes\cdots\sigma_0}_{d-s}\otimes\sigma_3\otimes\underbrace{\sigma_0\cdots\sigma_0}_{s-1},
\label{c1}
\end{aligned}
\eeqn
where $\sigma_{1,2,3}$ denote the three Pauli matrices and $\sigma_0$ is $2\times 2$ identity matrix. This case has been studied in Ref.~\onlinecite {Luo2023} and $H_d$ is the generalization of the 2D BBH model to arbitrary $d$ dimensions. Another case is that the matrices labeled by different index $s$ commute with each other, for example,
\beqn
\begin{aligned}
&\Gamma_{sa}^{(d)}=\underbrace{\sigma_0\otimes\cdots\sigma_0}_{d-s}\otimes\sigma_2\otimes\underbrace{\sigma_0\cdots\sigma_0}_{s-1},\\
&\Gamma_{sb}^{(d)}=\underbrace{\sigma_0\otimes\cdots\sigma_0}_{d-s}\otimes\sigma_1\otimes\underbrace{\sigma_0\cdots\sigma_0}_{s-1},\\
&C_{s}=i\Gamma_{sa}^{(d)}\Gamma_{sb}^{(d)}=\underbrace{\sigma_0\otimes\cdots\sigma_0}_{d-s}\otimes\sigma_3\otimes\underbrace{\sigma_0\cdots\sigma_0}_{s-1}.
\label{c2}
\end{aligned}
\eeqn
This case has been studied in Ref.~\onlinecite {liu2023} and $H_d$ is the generalization of the 2D SSH model to arbitrary $d$ dimension. 

For $d=2$, in addition to the two cases given by Eqs.~\eqref{c1} (2D BBH model) and \eqref{c2} (2D SSH model),   the condition that $C_1$ and $C_2$ commute with each other can be alternatively satisfied when
$[\Gamma_{1a}^{(2)},\Gamma_{2a,2b}^{(2)}]=0$ and $ \{\Gamma_{1b}^{(2)},\Gamma_{2a,2b}^{(2)}\}=0$, which is associated with the 2D crossed SSH model \cite{Luo2022}. In the rest of this paper, we construct the model Hamiltonians at $d=3$ and study the obtained higher-order topological phases.

\section{Model Construction at $d=3$}
\label{III}

\begin{figure*}
\centering
\includegraphics[width=7in]{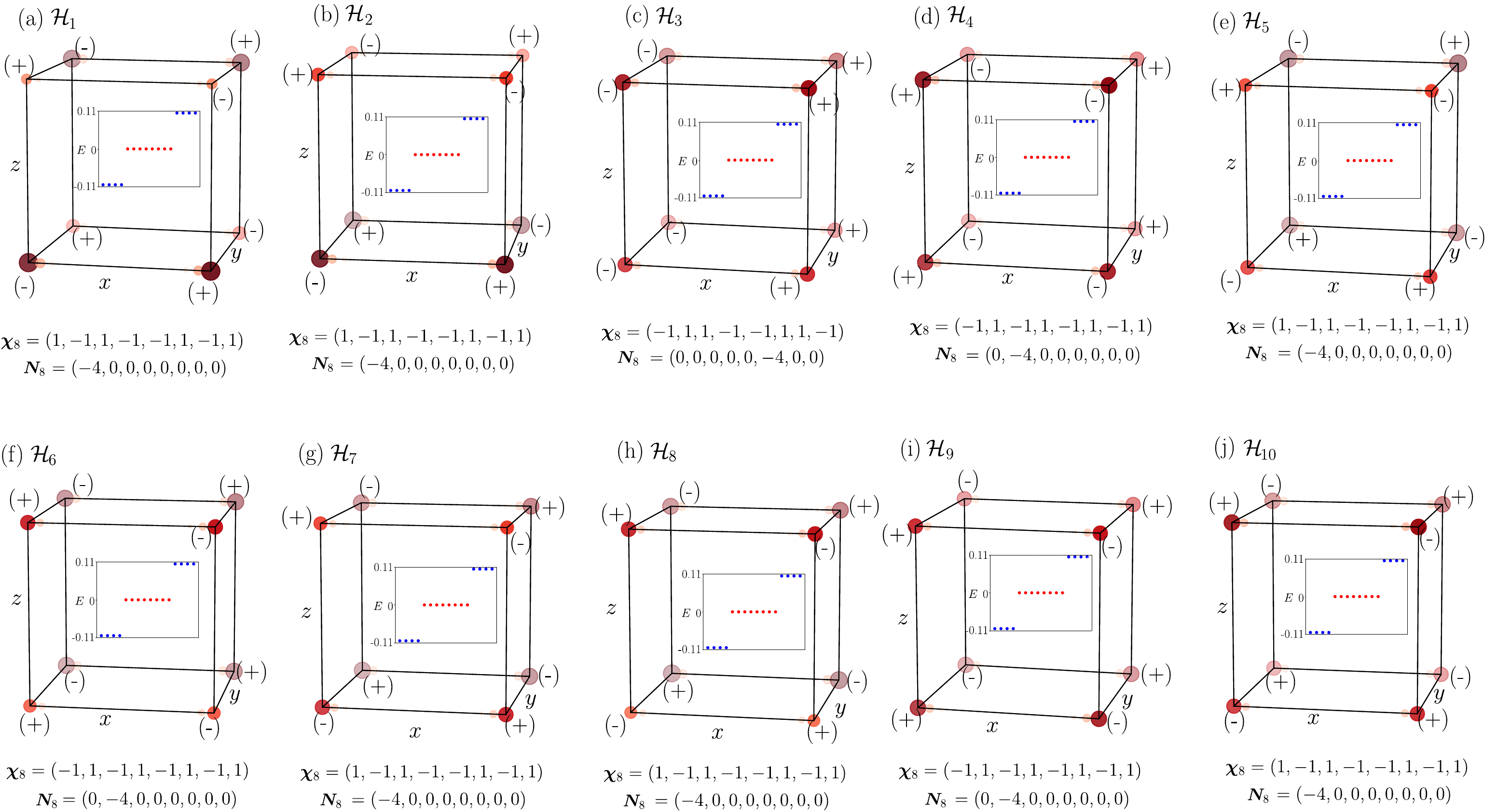}
\caption{(a)-(j) Spatial distribution of ZECSs of $\mathcal{H}_{1,\cdots,10}$. The insets in (a)-(j) plot the eigenenergies near zero under the open boundary conditions. The label ($z$) at each corner denotes the eigenvalue of ZECSs under the bulk chiral symmetry $\mathcal{C}$, with $z=\pm 1$. The vectors $\chi_8$ and $\bm{N}_8$ are used to describe and characterize different patterns of corner states. We take the common model parameters as $\lambda_1=0.2,\lambda_2=0.5,\lambda_3=2$, and $t_1=t_2=t_3=0.1$.  }
\label{Fig2}
\end{figure*}

For $d=3$,  we rewrite the Hamiltonian in Eq.~\eqref{h1} as
\beqn
\begin{aligned}
&\mathcal{H}(\bm k_3)=\sum_{s=1}^3h_s(k_s),\\
&h_s(k_s)=M_s(k_s)\Gamma_{sa}^{(3)}+\lambda_s\sin k_s\Gamma_{sb}^{(3)},
\label{S3D}
\end{aligned}
\eeqn
where $\{\Gamma_{sa}^{(3)},\Gamma_{sb}^{(3)}\}=0$ and $h_s$ respects the chiral symmetry $C_s=-i\Gamma_{sa}^{(3)}\Gamma_{sb}^{(3)}$. Generally, when $\{C_1,C_2,C_3\}$ commute with each other, the 3D Hamiltonian $\mathcal{H}$ hosts eight ZECSs described by the wave function
\beqn
|\Psi_{z_1,\cdots,z_3}(\bm r_3)\rangle=\mathcal{N}\prod_{s=1}^3f_{z_s}^s(r_s)|\psi_{(z_1,\cdots,z_3)}\rangle,
\label{SZECS}
\eeqn
with $z_s=\pm 1$ being the eigenvalue of $C_s$. We note that the localized position of the eight ZECSs in real space is completely determined by their  eigenvalues $(z_1,z_2,z_3)$.  To be specific, when $z_s=-1~(1)$, ZECS $\Psi_{z_1,z_2,z_3}(\bm r)$ is localized close to the corner where $r_s=-L_s/2~(L_s/2)$, with $s=1,2,3$. When $C_1\neq \pm C_2\neq \pm C_3$, the eight common eigenstates of $\{C_1,C_2, C_3\}$ are non-degenerate about the eigenvalues $(z_1,z_2,z_3)$. Therefore, there is one ZECS localized at each corner of the 3D cubic system.
  Once $C_s=C_{s^{\prime}}$ or $C_s=-C_{s^{\prime}}$ with $s\neq s^{\prime}$, the common eigenstates of $\{C_1,C_2, C_3\}$ must have degeneracies about the eigenvalues $(z_1,z_2,z_3)$ under the restriction $z_s=z_{s^{\prime}}$ or $z_s=-z_{s^{\prime}}$. These degeneracies imply that there are multiple ZECSs localized at a certain corner of the $3$D cubic lattice. In the following, we make an enumeration of $\mathcal{H}$ under the condition that there is one ZECS at each corner.

\begin{figure*}
\centering
\includegraphics[width=7in]{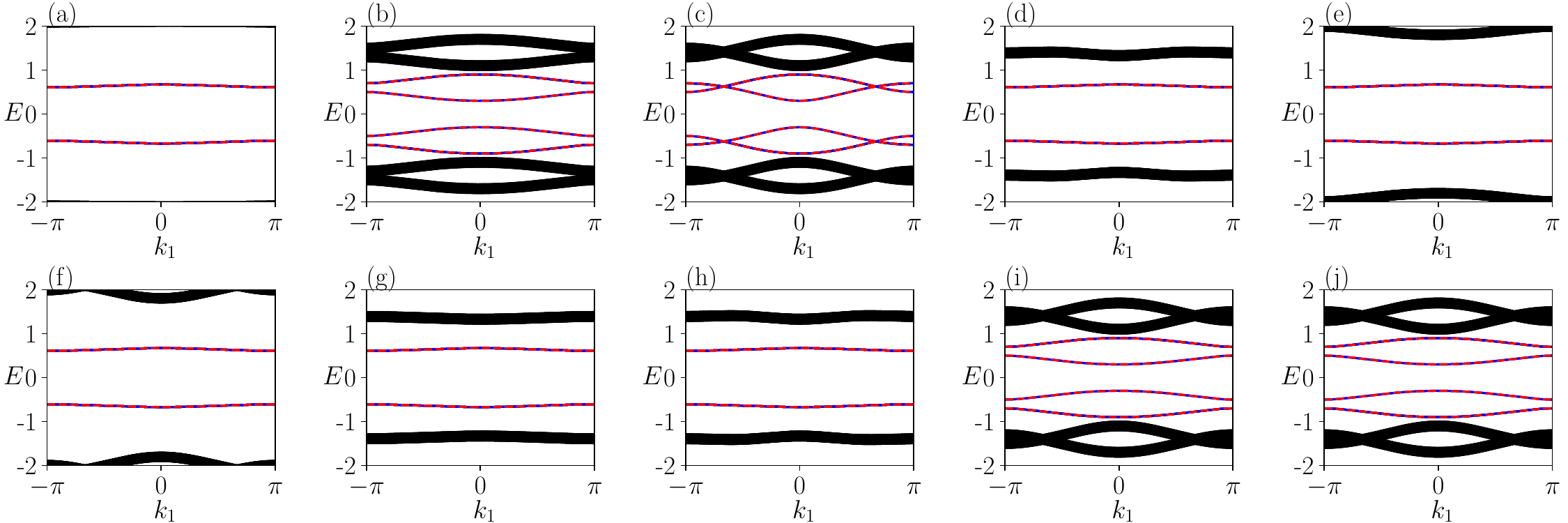}
\caption{Energy spectra of $\mathcal{H}_{1,\cdots,10}$ in a slab geometry with the open boundary conditions along the $r_3$ direction. In (a)-(j), we take $k_2=0$ and the in-gap blue bands correspond to the gapped surface states, of which the energy spectra can be accurately captured by the analytical results (dashed red bands).   The same model parameters are taken as those used in Fig.~\ref{Fig2}. }
\label{Fig3}
\end{figure*}

Informed by the 2D case \cite{Luo2022},  we consider three types of commutation or anti-commutation relations between matrices $\{\Gamma_{sa}^{(3)},\Gamma_{sb}^{(3)},\Gamma_{s^{\prime}a}^{(3)},\Gamma_{s^{\prime}b}^{(3)}\}$ under the condition that $\{C_1,C_2,C_3\}$ commute with each other,
\beqn
\begin{aligned}
&\text{i}: \{\Gamma_{sa}^{(3)},\Gamma_{s^{\prime}a,s^{\prime}b}^{(3)}\}=0, \{\Gamma_{sb}^{(3)},\Gamma_{s^{\prime}a,s^{\prime}b}^{(3)}\}=0,\\
&\text{ii}: [\Gamma_{sa}^{(3)},\Gamma_{s^{\prime}a,s^{\prime}b}^{(3)}]=0, [\Gamma_{sb}^{(3)},\Gamma_{s^{\prime}a,s^{\prime}b}^{(3)}]=0,\\
&\text{iii}: [\Gamma_{sa}^{(3)},\Gamma_{s^{\prime}a,s^{\prime}b}^{(3)}]=0, \{\Gamma_{sb}^{(3)},\Gamma_{s^{\prime}a,s^{\prime}b}^{(3)}\}=0,
\label{3c}
\end{aligned}
\eeqn
with $(s,s^{\prime})\in\{(1,2),(2,3),(1,3)\}$. We label the commutation or anti-commutation relations between matrices $\Gamma_{sa,sb}^{(3)}$ and $\Gamma_{s^{\prime}a,s^{\prime}b}^{(3)}$
as $(ss^{\prime}\text{w})$, with $\text{w}\in\{\text{i,ii,iii}\}$.  Therefore, there are 27 types of commutation or anti-commutation relations. We label these relations as $(12\text{w},23\text{w}^{\prime},13\text{w}^{\prime\prime})$, with $\text{w},\text{w}^{\prime},\text{w}^{\prime\prime}\in\{\text{i,ii,iii}\}$.
Note that these 27 types of relations are obtained when 
the equivalent status of permuting the direction index is not distinguished. After considering this equivalency, there are 10 types of inequivalent relations left, which are listed in Table~\ref{tab1}.

In Table~\ref{tab2}, we give the concrete representation of matrices $\{\Gamma_{1a,2a,3a}^{(3)},\Gamma_{1b,2b,3b}^{(3)}\}$ to satisfy the relations specified in Table~\ref{tab1}, which predict ten distinct models labeled as $\mathcal{H}_{1,\cdots,10}$. We find that $\mathcal{H}_1$ and $\mathcal{H}_2$ correspond to the well-studied 3D BBH \cite{Benalcazar2017,Benalcazar2017a} and 3D SSH \cite{liu2023} models, respectively,  while the other eight models have not been reported to our knowledge.  In Fig.~\ref{Fig1}, we present the schematic illustration of the lattice hoppings of $\mathcal{H}_{1,\cdots,10}$. In the Appendix~\ref{Appendixnd}, we present the details about the correspondence between these hopping patterns and the Bloch Hamiltonians.

We also present the bulk chiral symmetry $\mathcal{C}$ of $\mathcal{H}_{i}$ with $\{\mathcal{C},\mathcal{H}_i\}=0$ for $i=1,\cdots 10$ in Table~\ref{tab2}. We find that $\mathcal{C}$ can be expressed in terms of matrices $C_{1,2,3}$. Specifically, $\mathcal{C}=C_1C_2C_3$ for $\mathcal{H}_{1,2,5,7,8,10}$, $\mathcal{C}=C_1$ for $\mathcal{H}_3$, and $\mathcal{C}=C_1C_2$ for $\mathcal{H}_{4,6,9}$. Since ZECS $|\Psi_{z_1,\cdots,z_3}(\bm r_3)\rangle$ is the eigenstate of $C_s$ with eigenvalue $z_s$ for $s=1,2,3$, $|\Psi_{z_1,\cdots,z_3}(\bm r_3)\rangle$ is also the eigenstate of $\mathcal{C}$ with eigenvalue $z=z_1z_2z_3$ for $\mathcal{H}_{1,2,5,7,8,10}$, $z=z_1$ for $\mathcal{H}_3$, and $z=z_1z_2$ for  $\mathcal{H}_{4,6,9}$, which are associated with three types of chirality configurations in real space of the eight ZECSs, as shown in Fig.~\ref{Fig2}. With this property, a local perturbation $h_p$ preserving the chiral symmetry $\mathcal{C}$ with $\{h_p,\mathcal{C}\}=0$ can not remove the zero-energy state localized at a given corner because
\beqn
\begin{aligned}
&\langle\Psi_{z_1,z_2,z_3}|h_p|\Psi_{z_1,z_2,z_3}\rangle\\
&=z\langle\Psi_{z_1,z_2,z_3}|\mathcal{C}h_p+h_p\mathcal{C}|\Psi_{z_1,z_2,z_3}\rangle/2=0.
\end{aligned}
\eeqn
Thus, ZECSs are protected by the bulk chiral symmetry $\mathcal{C}$. For the same reasons, it can also be shown that the perturbation terms that anti-commute with one of the operators $\{C_1,C_2,C_3\}$ can not remove the ZECSs.  

While the Hamiltonians $\mathcal{H}_{i}$ for $i=1,\cdots 10$ always host ZECSs when $|t_s|<|\lambda_s|$ by construction, the TOTI phase further requires that the bulk, surface, and hinge states are all gapped, which is realized in certain but not all parameter regimes. For example, TOTIs featuring CZESs protected by an energy gap  
 can be realized in all the ten Hamiltonians under the parameters $\lambda_1=0.2,\lambda_2=0.5,\lambda_3=2$, and $t_1=t_2=t_3=0.1$, as shown in Fig.~\ref{Fig2}. In addition to the TOTI phase, the Hamiltonians can host other topological phases, which we discuss in Sec.\ref{VI}.

\section{Surface and hinge projection analyses}
\label{IV}

By a boundary projection analysis for the 3D BBH model \cite{Luo2023}, it was shown that all the 2D surfaces and 1D hinges can be described by the 2D BBH model and 1D SSH model, respectively. 
In this section, we focus on the TOTIs phase of the ten predicted models with gapped bulk, surface, and hinge states.
Similar to the 3D BBH model, we find that certain (but not all) surfaces and hinges of the systems described by $\mathcal{H}_{2,...,10}$ in the TOTI phase can also exhibit nontrivial second-order and first-order topology, respectively. We summarize the boundary topology for the ten models in Tabel \ref{tab3}. The Hamiltonians of the ten models are written in a unified manner as the summation of the extended SSH models along different directions, but the exact ways of stacking are different. Therefore, the ten models can have different boundary topologies as presented in Table \ref{tab3}. Nevertheless, all the ten models exhibit nontrivial surface topology by taking the open boundary condition along the $r_3$ direction, and nontriival hinge topology by taking the open boundary conditions along $r_2$ and $r_3$ directions, which we discuss in detail in the following.

\begin{table}[b]
\centering
\setlength\tabcolsep{2.2pt}
\renewcommand{\arraystretch}{2}
\caption{\textcolor{black}{A summary of the boundary topology for the ten models. The boundaries $\mathcal{X}$ are obtained by taking the open boundary conditions along the $\mathcal{X}$ direction, with $\mathcal{X}\in \{r_1,r_2,r_3,r_1r_2,r_1r_3,r_2r_3\}$. Symbols
$\checkmark$ and $\times$ denote nontrivial and trivial boundary topologies, respectively. }}
\begin{tabular}{|c|c|c|c|c|c|c|c|c|c|c|}
\hline
\diagbox[width=7em]{Boundaries}{Models}& $\mathcal{H}_1$& $\mathcal{H}_2$& $\mathcal{H}_3$ & $\mathcal{H}_4$ &$\mathcal{H}_5$  &$\mathcal{H}_6$& $\mathcal{H}_7$& $\mathcal{H}_8$& $\mathcal{H}_9$& $\mathcal{H}_{10}$  \\
\hline
$r_1$ & \checkmark & \checkmark &$\times$ &$\times$  &\checkmark &$\times$  &\checkmark &$\times$  &$\times$  &$\times$ \\
\hline
$ r_2$& \checkmark & \checkmark &$\times$  &\checkmark &\checkmark &\checkmark&\checkmark &$\times$ &\checkmark &$\times$  \\
\hline
 $r_3$& \checkmark & \checkmark &\checkmark &\checkmark &\checkmark &\checkmark &\checkmark &\checkmark &\checkmark &\checkmark \\
\hline
$r_1r_2$& \checkmark & \checkmark &$\times$ &$\times$  &\checkmark &$\times$  &\checkmark &$\times$  &$\times$  &$\times$ \\
\hline
$r_1r_3$&  \checkmark & \checkmark &$\times$ &\checkmark &\checkmark &\checkmark &\checkmark &\checkmark &\checkmark &\checkmark \\
\hline
$r_2r_3$&  \checkmark & \checkmark &\checkmark &\checkmark &\checkmark &\checkmark&\checkmark &\checkmark &\checkmark &\checkmark \\
\hline
\end{tabular}
\label{tab3}
\end{table}

\begin{figure*}
\centering
\includegraphics[width=7in]{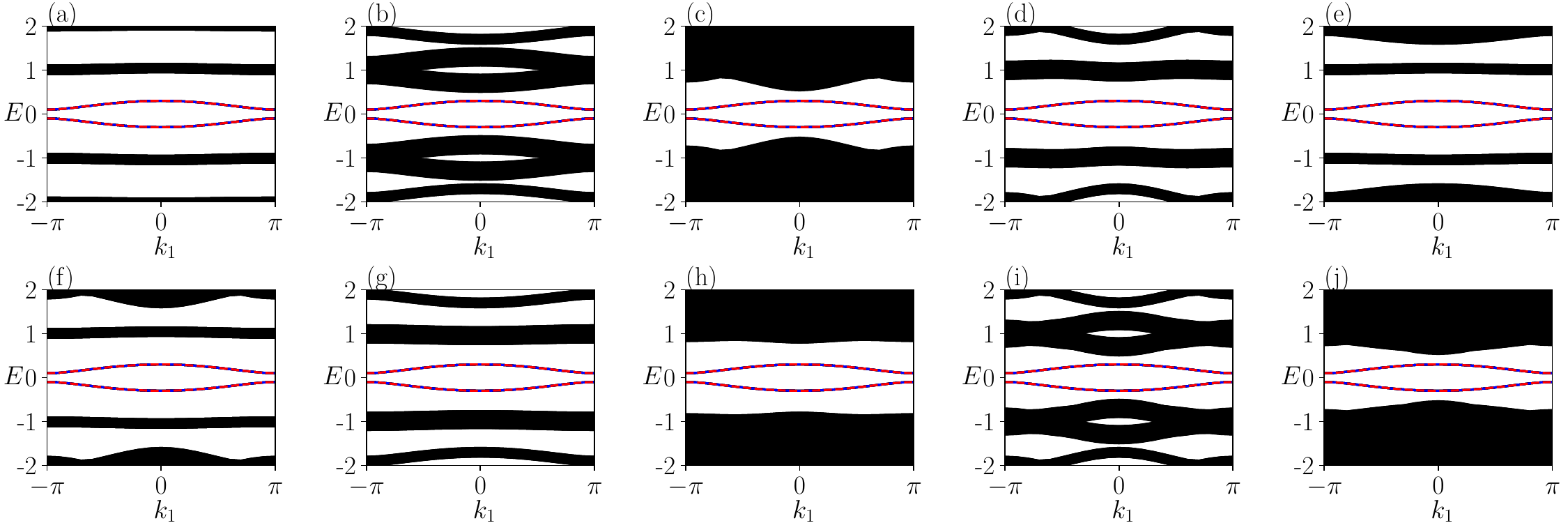}
\caption{(a)-(j) Energy spectra of $\mathcal{H}_{1,\cdots,10}$ in a wire geometry with the open boundary conditions along $r_2$ and $r_3$ directions. The in-gap blue bands correspond to the hinge states, of which the energy spectra can be accurately captured by the analytical result $E=\pm \sqrt{M_1^2+(\lambda_1\sin k_1)^2}$ (dashed red lines). The same model parameters are used as those in Fig.~\ref{Fig2}. 
}
\label{Fig4}
\end{figure*}

\subsection{Nontrivial surface topology}

In the TOTI phase, there are gapped surface states on the (001) surface for $\mathcal{H}_{1,\cdots,10}$ in a slab geometry, as shown in Fig. 3. Distinguished from the gapless boundary states of topological insulators,  these surface states are extended over the whole surface Brillouin zone and  can be exactly described by a lattice Hamiltonian. To derive the surface Hamiltonian, we decompose $\mathcal{H}_{i}$ as $\mathcal{H}_{i}(r_3,k_1,k_2)=h_3(r_3)+h_{12}(k_1,k_2)$, with $h_{12}=h_1+h_2$ and $i=1,\cdots,10$.  We note that $[C_3,h_{12}]=0$ for $\mathcal{H}_{1,\cdots,10}$ and the end zero-energy states of $h_3$ are the eigenstates of $C_3$. Thus, the surface states of $\mathcal{H}_{i}(r_3,k_1,k_2)$ are the common eigenstates of $C_3$ and $h_{12}$
\beqn
\begin{aligned}
&|\Phi_{z_3}(r_3,k_1,k_2)\rangle=f_{z_3}^{(3)}(r_3)P_{z_3}^{(3)}|\phi(k_1,k_2)\rangle,\\
&h_{12}(k_1,k_2)|\phi(k_1,k_2)\rangle=E(k_1,k_2)|\phi(k_1,k_2)\rangle,
\end{aligned}
\eeqn
where the surface projection operator is $P_{z_3}^{(3)}=(1+z_3C_3)/2$ and the surface states have the same energy spectra as those of $h_{12}$. For models $\mathcal{H}_{1,4,5,6,7,8}$,  $\mathcal{H}_{2,9,10}$, and $\mathcal{H}_{3}$, the energy spectra  of $h_{12}$, are respectively,
\beqn
\begin{aligned}
&E(k_1,k_2)=\pm \sqrt{\mathcal{E}_1^2+\mathcal{E}_2^2},\\
&E(k_1,k_2)=\pm \mathcal{E}_1\pm\mathcal{E}_2,\\
&E(k_1,k_2)=\pm \sqrt{(M_1\pm \mathcal{E}_2)^2+\lambda_2^2\sin k_2^2},
\end{aligned}
\eeqn
where $\mathcal{E}_s=\sqrt{M_1^2+\lambda_s^2\sin k_s^2}$ for $s=1,2$. As shown in Fig.~\ref{Fig3}, the surface energy spectra given by the analytical results perfectly agree with the numerical results.

The surface Hamiltonian can be extracted by projecting $\mathcal{H}_{i}$ onto the subspace defined by $P_{z_3}^{(3)}$.  For example, the surface Hamiltonian for the surface $r_3=-L_3/2$ can be written as
\beqn
\begin{aligned}
\tilde{h}_{12}(k_1,k_2)&=P_{-}^{(3)}\mathcal{H}_iP_{-}^{(3)}\\
&=\sum\limits_{s=1}^2M_s(k_s)\tilde{\Gamma}_{sa}^{(3)}+\lambda_s\sin k_s\tilde{\Gamma}_{sb}^s,
\end{aligned}
\eeqn
where we have defined $\tilde{\Gamma}_{sa,sb}^{(3)}=P_{-}^{(3)}\Gamma_{sa,sb}^{(3)}P_{-}^{(3)}$. Because  $[C_3,h_{12}]=0$,  matrices $\{\tilde{\Gamma}_{1a}^{(3)},\tilde{\Gamma}_{1b}^{(3)},\tilde{\Gamma}_{2a}^{(3)},\tilde{\Gamma}_{2b}^{(3)}\}$ obey the same commutation relations as $\{\Gamma_{1a}^{(3)},\Gamma_{1b}^{(3)},\Gamma_{2a}^{(3)},\Gamma_{2b}^{(3)}\}$.  Thus, once projecting onto the non-zero block of $P_{-}^{3}$, the $4\times 4$ surface Hamiltonian $\tilde{h}_{12}(k_1,k_2)$ are equivalent to the 2D BBH model, the 2D SSH model, and the 2D crossed SSH model for $\mathcal{H}_{1,4,5,6,7,8}$,  $\mathcal{H}_{2,9,10}$, and $\mathcal{H}_{3}$, respectively. These three models behave as second-order topological insulators under appropriate model parameters \cite{Luo2022}.

The existence of surface states on the (001) surface for $\mathcal{H}_{1,\cdots,10}$ can be attributed to the commutation relation $[C_3,h_{12}]=0$. It can be  verified that $[C_1,h_{23}]=0$ for $\mathcal{H}_{1,2,5,7}$ and $[C_2,h_{13}]=0$ for $\mathcal{H}_{1,2,4,5,6,7,9}$ with $h_{23}=h_2+h_3$ and $h_{13}=h_1+h_3$. Therefore, there are surface states localized at the (100) and (010) surfaces for $\mathcal{H}_{1,2,5,7}$ and $\mathcal{H}_{1,2,4,5,6,7,9}$, respectively, in a slab geometry. For the same reasons, these surface states are described by the 2D BBH model, the 2D SSH model, or the 2D crossed SSH model.

\begin{figure*}
\centering
\includegraphics[width=7in]{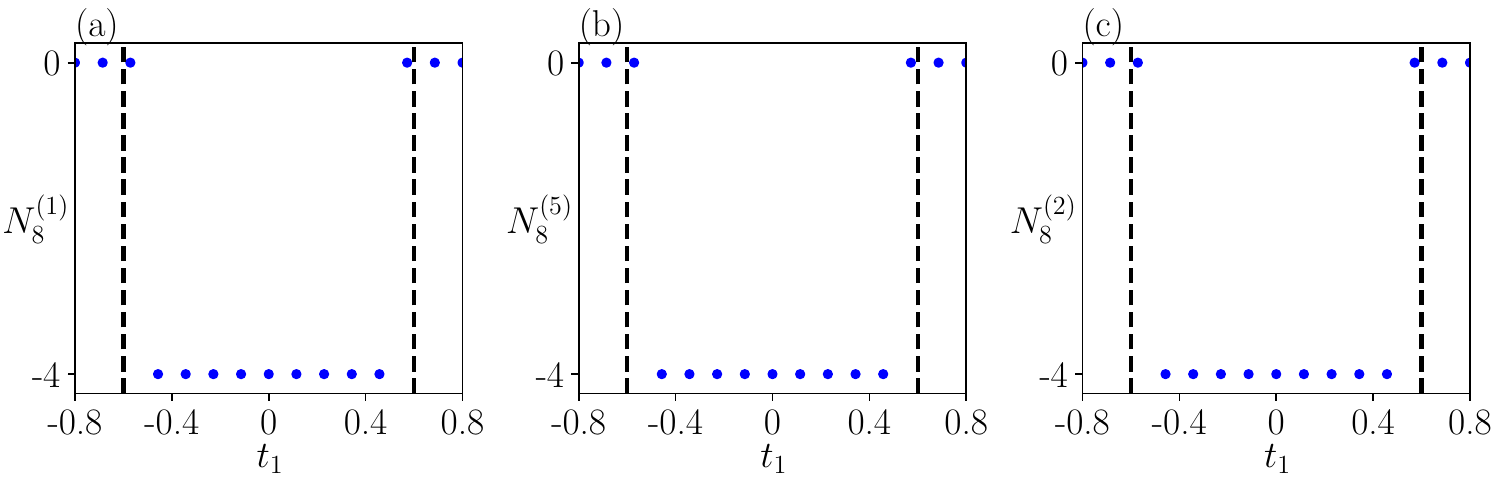}
\caption{Numerical results of $N_8^{(1,5,2)}$ as a functions of $t_1$ by setting $\lambda_1=0.6,\lambda_2=1.5,\lambda_3=3$. In the numerical calculation, the size of the systems is taken as $11\times 11\times 11$. 
The black dashed lines mark the expected topological phase transition at $|t_1|=|\lambda_1|$. 
For parameters used in this figure, the bulk and surface states are always gapped, and the hinge states are gapped except at the transition point. (a), (b), and (c) are calculated for the models $\mathcal{H}_{1}$, $\mathcal{H}_3$, and $\mathcal{H}_{4}$, respectively.
}
\label{Fig6}
\end{figure*}

\subsection{Nontrivial hinge topology}
Since the 2D BBH,  the 2D SSH, and the 2D crossed SSH model host gapped edge states \cite{Luo2022}, there are gapped hinge states for $\mathcal{H}_{1,\cdots,10}$ in a wire geometry with open boundary conditions along $r_2$ and $r_3$ conditions, as shown in Fig.~\ref{Fig4}. These gapped hinge states can also be exactly described by a lattice Hamiltonian. To derive the hinge Hamiltonian, we decompose $\mathcal{H}_i$ as
\beqn
\mathcal{H}_{i}(k_1,r_2,r_3)=h_1(k_1)+h_{23}(r_2,r_3), 
\eeqn
where $h_{23}=h_2+h_3$ and $i=1,\cdots,10$. 
With a similar ZECSs construction method presented in Sec.~\ref{III}, it can be shown that $h_{23}$ also hosts eight ZECSs, which are the common eigenstates of $C_2$ and $C_3$. Since $[C_{2,3},h_1]=0$,  the gapped hinge states are the common eigenstates of $C_{2,3}$ and $h_1$ 
\beqn
\begin{aligned}
&|\Phi_{z_2,z_3}(k_1,r_2,r_3)\rangle=\prod_{s=2}^3f_{z_s}^{(s)}P_{z_s}^{(s)}|\phi(k_1)\rangle,\\
&h_1(k_1)|\phi(k_1)\rangle= E(k_1)|\phi(k_1)\rangle,
\end{aligned}
\eeqn
where the projection operator is $P_{z_s}^{(s)}=(1+z_sC_s)/2$ and the hinge energy spectra are $E=\pm \mathcal{E}_1$,  which perfectly agree with the numerical results as shown in Fig.~\ref{Fig4}.

The hinge Hamiltonian can be extracted by projecting $\mathcal{H}_i$ onto the subspace defined by the hinge projection operator $P_{z_2z_3}=P_{z_2}^{(2)}P_{z_3}^{(3)}$. For example, the hinge Hamiltonian for  hinge $(r_2,r_3)=(-L_2/2,-L_3/2)$ can be written as
\beqn
\begin{aligned}
\tilde{h}_1(k_1)&=P_{--}\mathcal{H}_iP_{--}\\
&=M_1(k_1)\tilde{\Gamma}_{1a}^{(3)}+\lambda_1\sin k_1\tilde{\Gamma}_{1b}^{(3)},
\end{aligned}
\eeqn
where $\tilde{\Gamma}_{1a,1b}^{(3)}=P_{--}\Gamma_{1a,1b}^{(3)}P_{--}$. As $[P_{--}, \Gamma_{1a,1b}^{(3)}]=0$, we have $\{\tilde{\Gamma}_{1a}^{(3)},\tilde{\Gamma}_{1b}^{(3)}\}=0$. The non-zero block of $P_{--}$ is $2\times 2$ as $P_{--}$ is the product of two projection operators. Thus when projecting onto the non-zero block of $P_{--}$,  $\tilde{h}_1(k_1)$  exactly takes the form of a two-band SSH model, which is topologically nontrivial and characterized by the winding number $\tilde{\nu}_1=1$ defined by the occupied states of $\tilde{h}_1(k_1)$.

We note that the existence of gapped hinge states of $\mathcal{H}_{1,\cdots,10}$ along the $k_1$ direction can be traced to the commutation relation $[C_{2,3},h_1]=0$. It can be verified that $[C_{1,3},h_2]=0$ and $[C_{1,2},h_3]=0$ for $\mathcal{H}_{1,2,4,\cdots, 10}$ and $\mathcal{H}_{1,2,5,7}$, respectively. Thus, $\mathcal{H}_{1,2,4,\cdots, 10}$ and $\mathcal{H}_{1,2,5,7}$ in a wire geometry host gapped hinge states along the $k_2$ and $k_3$ direction, respectively, which are described by a two-band SSH model.

Since a corner of a system is the end of a hinge, the nontrivial hinge topology naturally gives rise to the corner states. Inversely, the existence of ZECSs is rooted in the nontrivial hinge topology. 
Similarly, the existence of ZECSs can be attributed to the nontrivial surface topology.
The cascaded boundary nontrivial topology reveals the dimensional hierarchies of the systems.  

\section{Topological characterization }
\label{V}
In our recent work of Ref.~\onlinecite{Jiazheng2024}, a comprehensive theoretical framework has been established to universally characterize the chiral symmetric systems by a series of Bott indices. Generally, $(m-1)$ Bott indices are needed to fully characterize the corner states in a system with $m$ corners. This topological characterization can capture the different patterns of corner states described by the vector
\begin{eqnarray}
\boldsymbol{\chi}_{m}=\left(n_1^{-}-n_1^{+},\dots,n_m^{-}-n_m^{+}\right),
\end{eqnarray}
where $n_i^{-(+)}$ denotes the number of corner states that are localized at the $i$th corner with eigenvalue $-1$ ($1$) of chiral symmetry $\mathcal{C}$ and $m=8$ for the cubic-shaped systems. As shown in Fig.~\ref{Fig2}, the pattern of corner states of $\mathcal{H}_{1,2,5,7,8,10}$, $\mathcal{H}_{3}$, and $\mathcal{H}_{4,6,9}$ are described by different $\bm{\chi}_8$ vectors because of the different chirality configurations of corner states. Following the framework established in Ref.~\onlinecite{Jiazheng2024}, we use the Bott index vector $\bm{N}_8=({N}_8^{(1)},\cdots,{N}_8^{(7)})$ to characterize the vector $\bm{\chi}_8$ for each model and ${N}_{8}^{(1,\cdots,7)}$ are formulated as follows.

Under the eigenbasis of $\mathcal{C}$ where $\mathcal{C}=\sigma_z$, $\mathcal{H}_{1,\cdots,10}$ can be expressed in an off-diagonal form
\begin{eqnarray}
\mathcal{H}_i=\begin{pmatrix} 0 & h\\
h^{\dagger}& 0 
\end{pmatrix}.
\label{eqh}
\end{eqnarray}
By the singular value decomposition $h=U_A\Sigma U_B^{\dagger}$, we can define the Bott index in the open boundary condition as \cite{Jiazheng2024} 
\begin{eqnarray}
N_8^{(i)}&=&\text{Bott}(m_i,q)\equiv\frac{1}{2\pi i}\text{Tr}\text{log}(m_iqm_i^{\dagger}q^{\dagger})
\end{eqnarray}
where $m_i=e^{2\pi if_i(\bm r)}$ is an unitary matrix generated by polynomial $f_i(\bm r)$ and $q=U_AU_B^{\dagger}$. The polynomials can be chosen as \cite{Jiazheng2024}
\begin{eqnarray}
&&f_1(\bm r)=4r_1r_2r_3/L_1L_2L_3,\nonumber\\
&&f_2(\bm r)=2r_1r_2/L_1L_2,\nonumber\\
&&f_3(\bm r)=2r_1r_3/L_1L_3,\nonumber\\
&&f_4(\bm r)=2r_2r_3/L_1L_3,\nonumber\\
&&f_5(\bm r)=r_1/L_1,\nonumber\\
&&f_6(\bm r)=r_2/L_2,\nonumber\\
&&f_7(\bm r)=r_3/L_3.
\end{eqnarray}
Then the vector $\bm{\chi}_8$ can be characterized as
\begin{eqnarray}
\boldsymbol{\chi}_{8}=\mathcal{M}^{-1}\cdot\left({{N}}_8^{(1)},\dots,{{N}}_8^{(7)},0\right)^{\mathrm{T}},
\label{cc}
\end{eqnarray}
where $\mathcal{M}_{ij}=\operatorname{sign}\left({f}_{i}(\boldsymbol{x}_{j})\right)/2$ for $1\le i \le 7$, $\mathcal{M}_{8j}=1/2$ with $\bm{x}_j$ being the position of $j$th corner, and  $\mathcal{M}$ is given by
\begin{equation}
\mathcal{M}=\frac{1}{2}\left(
\begin{array}{cccccccccccc}
 -1 & 1 & -1 & 1 & 1 & -1 & 1 & -1 \\
 1 & -1 & 1 & -1 & 1 & -1 & 1 & -1  \\
 -1 & 1 & 1 & -1 & 1 & -1 & -1 & 1 \\
 -1 & -1 & 1 & 1 & 1 & 1 & -1 & -1  \\
 1 & -1 & -1 & 1 & 1 & -1 & -1 & 1  \\
 1 & 1 & -1 & -1 & 1 & 1 & -1 & -1  \\
 -1 & -1 & -1 & -1 & 1 & 1 & 1 & 1  \\
 1 & 1 & 1 & 1 & 1 & 1 & 1 & 1  \\
\end{array}
\right).
\end{equation}

In Fig.~\ref{Fig2}, we present the numerical results of $\bm{N}_8$ for the ten models in the TOTIs phase, which is consistent with Eq.~\eqref{cc}. We note that for $\mathcal{H}_{1,2,5,7,8,10}$, $\mathcal{H}_3$, and $\mathcal{H}_{4,6,9}$, only the Bott indices ${N}_{8}^{(1)}$, ${N}_{8}^{(5)}$, and ${N}_{8}^{(2)}$ are, respectively, nonzero. Thus, we characterize the topological phase transition of $\mathcal{H}_{1,2,5,7,8,10}$, $\mathcal{H}_3$, and $\mathcal{H}_{4,6,9}$ by ${N}_{8}^{(1)}$, ${N}_{8}^{(5)}$, and ${N}_{8}^{(2)}$, respectively, as shown in  Fig.~\ref{Fig6}.

\begin{figure}
\centering
\includegraphics[width=3.4in]{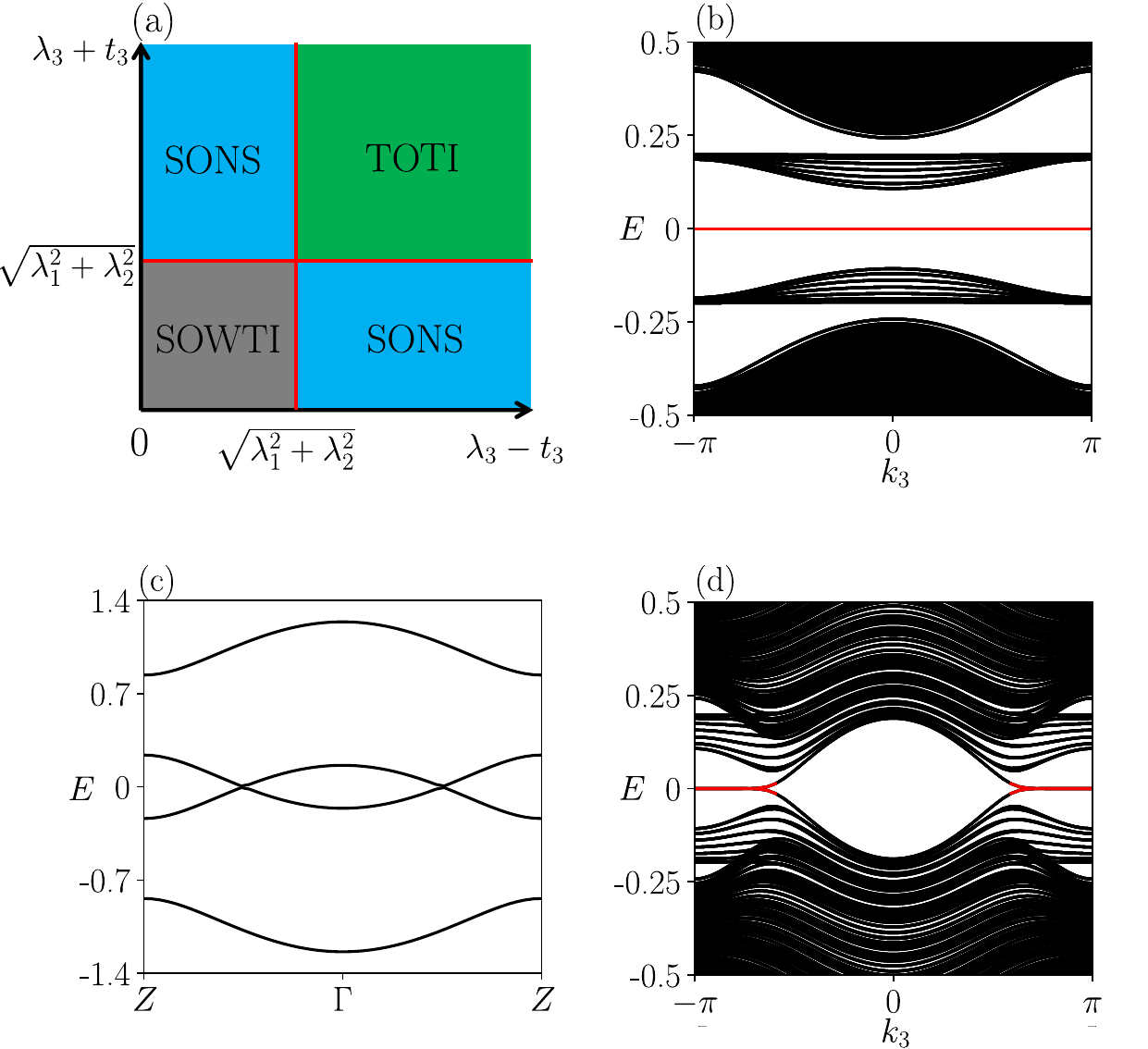}
\caption{(a) The phase diagram of $\mathcal{H}_8$ as a function of $\lambda_3-t_3$ and $\lambda_3+t_3$ by setting $t_1=t_2=0$ and $\lambda_3>|t_3|>0$. The phase transitions, represented by the red lines, divide the system into second-order weak topological insulators (SOWTI), second-order nodal semimetal (SONS), and TOTI phases, corresponding to the gray, blue, and green regions, respectively. (b) The energy spectrum of the SOWTI phase in a wire geometry with the open boundary conditions along $k_1$ and $k_2$ directions. (c)
The bulk energy spectrum along $Z-\Gamma-Z$ for the SONS phase. (d) The energy spectrum of the SONS phase in a wire geometry with the open boundary conditions along $k_1$ and $k_2$ directions. In (b), we take  $\lambda_1=0.2,\lambda_2=0.5,\lambda_3=0.2,t_3=0.1$. In (c) and (d), we take  $\lambda_1=t_3=0.2,\lambda_2=\lambda_3=0.5$. 
}
\label{Fig5}
\end{figure}

\section{Phase diagram of $\mathcal{H}_8$}
\label{VI}
The bulk energy spectra of the 3D BBH model are always gapped unless $|t_{1,2,3}|=|\lambda_{1,2,3}|$, while the gapped conditions for the bulk energy spectra of $\mathcal{H}_{2,\cdots,10}$ are generally complicated (see Appendix~\ref{appendixD}). 
Here we take $\mathcal{H}_8$ as an example and present its phase diagram in the parameter regions $|\lambda_{1,2,3}|>|t_{1,2,3}|$ where ZECSs exist.

The bulk energy spectrum of $\mathcal{H}_8$ is 
\beqn
E_{8}=\pm \sqrt{\mathcal{E}_1^2+\mathcal{E}_2^2+\mathcal{E}_3^2\pm 2\mathcal{E}_3\sqrt{M_1^2+M_2^2}},
\eeqn
where $\mathcal{E}_{3}=\sqrt{M_3^2+(\lambda_3\sin k_3)^2}$. 
The system is gapped when $|\lambda_3|-|t_3|>\sqrt{\sum_{s=1,2}(|\lambda_s|+|t_s|)^2}$ or $|\lambda_3|+|t_3|<\sqrt{\sum_{s=1,2}(|\lambda_s|-|t_s|)^2}$. For the former case, the system behaves as a third-order topological insulator featuring ZECSs [Fig.~\ref{Fig2}(i)]. For the latter case, the system can be viewed as a stacking of second-order topological insulators with a given $k_3$. This phase features hinge flat bands over the whole hinge Brillouin zone, as shown in Fig.~\ref{Fig5}(b). We dub this phase as a weak second-order topological insulator. For parameters outside of the above two cases,
the bulk spectrum is closed at $(\alpha_1\pi,\alpha_2\pi,\pm Q_{\alpha_1,\alpha_2})$, where $\alpha_{1,2}=0, 1$ and
\beqn
\cos Q_{\alpha_1,\alpha_2}=\frac{-(t_3+\lambda_3)^2+\sum_{s=1,2}(t_s+(-1)^{\alpha_s}\lambda_s)^2}
{2t_3\lambda_3}. \nonumber\\
\eeqn 
In this third phase, the system behaves as a second-order nodal semimetal \cite{Lin2018,Wang2020,Ghorashi2020,Wang2022} which is characterized by bulk nodal points [Fig.~\ref{Fig5}(c)] connected by hinge flat bands [Fig.~\ref{Fig5}(d)].  The phase diagram encompassing these three phases is shown in Fig.~\ref{Fig5}(a). We note that in the second-order nodal semimetal phases, ZECSs coexist with the zero-energy bulk states and hinge states.  Therefore, $\mathcal{H}_8$ provides an example of bound states in the continuum \cite{Benalcazar2020}. A similar phenomenon can occur when the bulk states of $\mathcal{H}_{3,4,5}$ and $\mathcal{H}_{2,7,9}$
behave as a nodal line semimetal and metal, respectively. Therefore, it is also interesting to explore the phase diagram of the other models.

\section{Discussion and Conclusion}
\label{VII}
In summary, we construct a family of chiral symmetry-protected third-order topological insulators as 3D generalizations of the 1D SSH model. Our approach is guided by the analysis of corner states and an enumeration of the model Hamiltonians utilizing generalized Pauli matrices, which generate ten distinct models, including the 3D BBH model. 
While these ten models have a unified expression in terms of the extended SSH model, they have different matrix forms (Table~\ref{tab2}), real-space hopping patterns (Fig.~\ref{Fig1}), and chirality configurations for corner states (Fig.~\ref{Fig2}). We emphasize that the certain hinge of the ten models is topologically nontrivial and exactly described by the two-band SSH model. This enables the realization of a family of boundary topological insulators \cite{2024BTIluo}.

Our study broadens the range of theoretical model Hamiltonians for chiral symmetry-protected TOTIs, potentially stimulating further theoretical and experimental research.  It is challenging to realize the predicted models in solid state systems. Given the experimental realization of the 3D BBH model and TOTIs across various metamaterial platforms, such as acoustic \cite{XueHaoran2019, Ni2020, Xue2020}, photonic \cite{2024wang}, and electrical circuit systems \cite{Bao2019, Liu2020-le}, we propose that these metamaterials also hold great promise for simulating other predicted models. Moreover, we anticipate that the family of models we introduce could facilitate the application of higher-order topological phases in these systems, enabling advancements in areas such as energy harvesting \cite{Rainbow2024}, low-threshold lasing \cite{JinqiWu2021}, and high-quality factor nanocavities \cite{Yasutomo2019}.
While our primary focus is on 3D systems, the methodology can be generalized to 4D, providing potential examples for investigating fourth-order topological phases that could also be effectively simulated using electric circuit systems \cite{Zhang2020} and synthetic dimensions for photonics \cite{Yuan18}.

It is known that the SSH model and Kitaev chain have the same Hamiltonian description and are distinguished by the chosen basis. If we consider $h_s$ in Eq.~\eqref{h1} to be multiple copies of the Kitaev chain and written in the Majorana basis, the Hamiltonians $\mathcal{H}_{1,\cdots,10}$ describe the third-order topological superconductors featuring Majorana corner states. In this case, the unit cell plotted in Fig.~\ref{Fig1} is composed of eight  Majorana fermions, and each corner of the cubic system can host one isolated Majorana zero mode. Thus, our theory can also be applied to construct a class of higher-order topological superconductors.

\begin{appendix}

\section{ Generalized Pauli matrices}
\label{appendixA}

 We briefly review the generalized Pauli matrices. The higher dimensional generalization of Pauli matrices can be expressed by the direct product of $d$ copies of Pauli matrices and $2\times 2$ identity matrix $\Gamma_{i,\cdots,j\cdots,k}^{(d)}=\sigma_i\cdots\sigma_j\cdots\sigma_k$, where $i,j,k=0,1,2,3$,  $\sigma_{1,2,3}$ are the three Pauli matrices, and $\sigma_0$ is the $2\times 2$ identity matrix. In total, there are $4^d$ matrices with dimension $2^d\times 2^d$, which satisfy $(\Gamma_{i,\cdots,j\cdots,k}^{(d)})^2=1$ and $(\Gamma_{i,\cdots,j\cdots,k}^{(d)})^{\dagger}=\Gamma_{i,\cdots,j\cdots,k}^{(d)}$. These $4^d$ matrices form a group which can be generated by the $(2d+1)$  anti-commuting Gamma matrices $\Gamma^{(d)}_{1,2,\cdots,2d+1}$. These Gamma matrices form the Clifford algebra $\{\Gamma_{j}^{(d)},\Gamma_{j^{\prime}}^{(d)}\}=2\delta_{jj^{\prime}}$ for $j,j^{\prime}=1,\cdots,2d+1$. When $d$=1, $\Gamma_{1,2,3}^{(1)}$ correspond to the three Pauli matrices $\sigma_{1,2,3}$. When $d$=2, $\Gamma_{1,2,3,4,5}^{(2)}$ correspond to the five anti-commuting Dirac matrices, which can be chosen as
\beqn
\begin{aligned}
&\gamma_{1,2,3}^{(2)}=\sigma_3\otimes \sigma_{1,2,3}, \gamma_4^{(2)}=\sigma_1\otimes \sigma_{0},\\
&\gamma_5^{(2)}=-\gamma_1^{(2)}\gamma_2^{(2)}\gamma_3^{(2)}\gamma_4^{(2)}=\sigma_2\otimes \sigma_0.
\end{aligned}
\eeqn
The anti-commuting $2^d\times 2^d$ Gamma matrices $\Gamma^{(2d)}_{1,2,\cdots,2d+1}$ can be generically obtained according to the iteration relation
\beqn
\begin{aligned}
&\Gamma_{1,2,\cdots,2d-1}^{(d)}=\sigma_3\otimes \gamma_{1,2,\cdots,2d-1}^{(d-1)},\\
& \Gamma_{2d}^{(d)}=\sigma_1\otimes I, \Gamma_{2d+1}^{(d)}=\sigma_2\otimes I,
\end{aligned}
\eeqn
where $ I$ denotes the $2^{d-1}\times 2^{d-1}$ identity matrix.

The products of the anti-commuting Gamma matrices generate the other generalized Pauli matrices. For example,  in addition to the five anti-commuting Dirac matrices $\Gamma_{1,2,3,4,5}^{(2)}$, the remaining generalized Pauli matrices for $d=2$ can be defined by $i\Gamma_{m}^{(2)}\Gamma_n^{(2)}$, with $m,n=1,2,3,4,5$. When $d=3$, in addition to the seven anti-commuting Gamma matrices $\Gamma_{1,\cdots,7}^{(3)}$, the other $8\times 8$ matrices can be generated by the successive products of two or three anti-commuting Gamma matrices, $i\Gamma_{m}^{(3)}\Gamma_n^{(3)}$ or $i\Gamma_m^{(3)}\Gamma_n^{(3)}\Gamma_l^{(3)}$, with $m,n,l=1,\cdots,7$.

\section{End states solution of the 1D extended SSH model}
\label{appendixB}
We present the solution for the end
zero-energy states (EZESs) of the 1D extended SSH model. The 1D extended SSH model can be written as
\beqn
h(k)=(t+\lambda\cos k)\Gamma_{a}^{(d)}+\lambda\sin k\Gamma_{b}^{(d)},
\label{II1}
\eeqn
where we omit the dimension label $s$ for convenience.
Expanding $h$ at $k=0$ (supposing that the energy spectrum of $h$ takes minimum value at $k=0$ ) to the second order of $k$ and replacing $k\rightarrow -i\partial_{r}$, we have
\beqn
h({r})=(m+\lambda/2\partial_{r}^2)\Gamma_{a}^{(d)}-i\lambda\partial_{r}\Gamma_{b}^{(d)},
\eeqn
with $m=t+\lambda$. Considering the semi-infinite system ($r>-L/2$), we solve the equation $h(r)|\Phi_{\alpha}(r)\rangle=0$, which yields
\beqn
(m+\lambda/2\partial_{r}^2)\Gamma_{a}^{(d)}|\Phi_{\alpha}(r)\rangle-i\lambda\partial_{r}\Gamma_{b}^{(d)}|\Phi_{\alpha}(r)\rangle=0.
\label{S1}
\eeqn
Multiplying both sides by $\Gamma_{a}^{(d)}$ in Eq.~\eqref{S1} yields
\beqn
(m+\lambda/2\partial_{r}^2)|\Phi_{\alpha}(r)\rangle=\lambda\partial_{r}C|\Phi_{\alpha}(r)\rangle,
\label{eq}
\eeqn
where $C=i\Gamma_a^{(d)}\Gamma_{b}^{(d)}$.
Therefore, state $|\Phi_{\alpha}(r)\rangle$ is the eigenstate of chiral symmetry $C$ and is labeled by the chiral symmetry eigenvalue $z=\pm 1$. We set the trial wave function $|\Phi_{z}(r)\rangle=e^{\xi_{z} r}|\psi_{z}\rangle$, where $\xi_{z}$ is a complex number, and spinor $|\psi_{z}\rangle$ satisfies $C|\psi_{z}\rangle=z|\psi_{z}\rangle$. Inserting this ansatz solution into Eq.~\eqref{eq}, we have
\beqn
\lambda/2\xi_{z}^2-z\lambda\xi+m=0,
\eeqn
which gives two roots $ \xi_{z}^{1,2}=\frac{z\lambda \pm \sqrt{\lambda^2-2m\lambda}}{\lambda}$. When $|t|<|\lambda|$, the real part of $ \xi_{z}^{1,2}$ are negative and positive when $z=-1$ and $z=1$, respectively. Under the boundary conditions $|\Phi_z(-L/2)\rangle=|\Phi_z(\infty)\rangle=0$, the wave function of the EZESs can be written as
\beqn
\begin{aligned}
&|\Phi_{-}(r)\rangle=f_{-}(r)|\psi_{-}\rangle,\\
& f_{-}(r)=\mathcal{N}_{-}(e^{\xi_{-}^{1}r}-e^{\xi_{-}^{2}r}),
\end{aligned}
\eeqn
where $\mathcal{N}_{-}$ is the normalization factor. Similarly, if we consider the semi-system $r<L/2$, the EZESs localized close to $r=L/2$ is the eigenstate of $C$ with eigenvalue $z=1$. Therefore, for a finite system with length $L$, the EZESs localized close to the end $r=-L/2$ and $r=L/2$ are the eigenstates of $C$, with eigenvalue $z=-1$ and $z=1$, respectively. Similarly, the wave function of ZECSs localized at $r=L/2$ can be written as
\beqn
\begin{aligned}
&|\Phi_{+}(r)\rangle=f_{+}(r)|\psi_{-}\rangle,\\
&f_{+}(r)=\mathcal{N}_{+}(e^{\xi_{+}^1(r-L)}-e^{\xi_{+}^2(r-L)}),
\end{aligned}
\eeqn
where $\mathcal{N}_{+}$ is the normalization factor.

\section{Correspondence between Bloch Hamiltonians and lattice hopping patterns}
\label{Appendixnd}
The Bloch Hamiltonian of the 3D BBH model can be written as
$$
\begin{aligned}
\mathcal{H}\left(\boldsymbol{k}_3\right) & =\sum_{s=1}^3 h_s\left(k_s\right), \\
h_s\left(k_s\right) & =M_s\left(k_s\right) \Gamma_{s a}^{(3)}+\lambda_s \sin k_s \Gamma_{s b}^{(3)},
\end{aligned}
$$
where $M_s\left(k_s\right)$ is $t_s+\lambda_s \cos k_s$ with $t_s$ and $\lambda_s$ being the model parameters. Matrices $\Gamma_{s a}^{(3)}$ and $\Gamma_{s b}^{(3)}$ are the $8 \times 8$ generalized Pauli matrix. In Table.~\ref{tab2}, we choose
\beqn
\begin{aligned}
\Gamma_{1 a}^{(3)}=\Sigma_{331}, 
\Gamma_{1 b}^{(3)}=\Sigma_{332}, \\
\Gamma_{2 a}^{(3)}=\Sigma_{310},
\Gamma_{2 b}^{(3)}=\Sigma_{320}, \\
\Gamma_{3 a}^{(3)}=\Sigma_{100}, 
\Gamma_{3 b}^{(3)}=\Sigma_{200},
\end{aligned}
\eeqn
where matrix $\Sigma_{i j k}$ is $\Sigma_{i j k}=\sigma_i \otimes \sigma_j \otimes \sigma_k$ with $i, j, k=0,1,2,3$. For the lattice model, the onsite term of $\mathcal{H}$ is
\beqn
\begin{aligned}
O&=t_1\Sigma_{331}+t_2\Sigma_{310}+t_3\Sigma_{100}\\
&=\left(\begin{matrix}
0& t_1&t_2&0&t_3&0&0&0\\
t_1&0&0&t_2&0&t_3&0&0\\
t_2&0&0&-t_1&0&0&t_3&0\\
0&t_2&-t_1&0&0&0&0&t_3\\
t_3&0&0&0&0&-t_1&-t_2&0\\
0&t_3&0&0&-t_1&0&0&-t_2\\
0&0&t_3&0&-t_2&0&0&t_1\\
0&0&0&t_3&0&-t_2&t_1&0
\end{matrix}\right).
\end{aligned}
\eeqn
The nonzero matrix element of $O$ implies there is hopping between the corresponding orbitals. For example, $O_{12}=t_1$ implies that there is hopping between the first and second orbitals. Thus, we can obtain the intracellular hopping patterns of the 3D BBH model, as shown in as shown in the inset of Fig.~\ref{Fig1}(a).

By the Fourier transformation, we can obtain the hopping matrix element along the $r_1, r_2$, and $r_3$ directions, respectively,
\beqn
\begin{aligned}
&T_1=\left(\begin{matrix}
0& \lambda_1&0&0&0&0&0&0\\
\lambda_1&0&0&0&0&0&0&0\\
0&0&0&-\lambda_1&0&0&0&0\\
0&0&-\lambda_1&0&0&0&0&0\\
0&0&0&0&0&-\lambda_1&0&0\\
0&0&0&0&-\lambda_1&0&0&0\\
0&0&0&0&0&0&0&\lambda_1\\
0&0&0&0&0&0&\lambda_1&0
\end{matrix}\right),
\end{aligned}
\eeqn
\beqn
\begin{aligned}
&T_2=\left(\begin{matrix}
0& 0&\lambda_2&0&0&0&0&0\\
0&0&0&\lambda_2&0&0&0&0\\
\lambda_2&0&0&0&0&0&0&0\\
0&\lambda_2&0&0&0&0&0&0\\
0&0&0&0&0&0&-\lambda_2&0\\
0&0&0&0&0&0&0&-\lambda_2\\
0&0&0&0&-\lambda_2&0&0&0\\
0&0&0&0&0&-\lambda_2&0&0
\end{matrix}\right),
\end{aligned}
\eeqn
\beqn
\begin{aligned}
&T_3=\left(\begin{matrix}
0& 0&0&0&\lambda_3&0&0&0\\
0&0&0&0&0&\lambda_3&0&0\\
0&0&0&0&0&0&\lambda_3&0\\
0&0&0&0&0&0&0&\lambda_3\\
\lambda_3&0&0&0&0&0&0&0\\
0&\lambda_3&0&0&0&0&0&0\\
0&0&\lambda_3&0&0&0&0&0\\
0&0&0&\lambda_3&0&0&0&0
\end{matrix}\right).
\end{aligned}
\eeqn
The nonzero matrix element of $T_{1,2,3}$ is associated with the hopping between different orbitals. Therefore, we can obtain the intercellular hopping patterns of the 3D BBH model, as shown in Fig.~\ref{Fig1}(a). The intracellular and intercellular hopping patterns of the other models can be obtained in a similar way.

\section{Bulk energy spectra }
\label{appendixD}
We give the analytical expression of the bulk spectra of $\mathcal{H}_{1,\cdots,9}$ while there is no analytical expression of the bulk energy spectra of $\mathcal{H}_{10}$. The bulk energy spectra of $\mathcal{H}_{1,\cdots,9}$ are
\beqn
&&{E}_{1}=\pm \sqrt{\mathcal{E}_1^2+\mathcal{E}_2^2+\mathcal{E}_3^2},\nonumber\\
&&{E}_{2}=\pm \mathcal{E}_1\pm \mathcal{E}_2\pm \mathcal{E}_3,\nonumber\\
&&E_3=\pm \sqrt{(M_1\pm \sqrt{(M_2\pm \mathcal{E}_3)^2+\lambda_2^2\sin k_2^2})^2+\lambda_1^2\sin k_1^2},\nonumber\\
&&E_{4}=\pm \sqrt{\mathcal{E}_1^2+\mathcal{E}_2^2+\mathcal{E}_3^2\pm 2\sqrt{(M_1^2+\mathcal{E}_2^2)\mathcal{E}_3^2}},\nonumber\\
&&E_{5}=\pm \sqrt{(\mathcal{E}_1\pm \mathcal{E}_3)^2+\mathcal{E}_2^2},\nonumber\\
&&E_{6}=\pm \sqrt{\mathcal{E}_1^2+\mathcal{E}_2^2+\mathcal{E}_3^2\pm 2M_1\mathcal{E}_3},\nonumber\\
&&E_{7}=\pm \sqrt{\mathcal{E}_1^2+\mathcal{E}_2^2}\pm E_3,\nonumber\\
&&E_{8}=\pm \sqrt{\mathcal{E}_1^2+\mathcal{E}_2^2+\mathcal{E}_3^2\pm 2\mathcal{E}_3\sqrt{M_1^2+M_2^2}},\nonumber\\
&&E_{9}=\pm \sqrt{\mathcal{E}_1^2+\mathcal{E}_3^2\pm 2M_1\mathcal{E}_3}\pm \mathcal{E}_2,
\eeqn
where $\mathcal{E}_{s}=\sqrt{M_s^2+(\lambda_s\sin k_s)^2}$ with $s=1,2,3$ and $E_{i}$ denotes the bulk energy spectra of $\mathcal{H}_{i}$.  The gapless conditions of $\mathcal{H}_{1,\cdots,9}$ at half filling are 
\beqn
&&E_1=0\Longrightarrow \mathcal{E}_1=\mathcal{E}_2=\mathcal{E}_3=0,\nonumber\\
&&E_2=0\Longrightarrow \mathcal{E}_s \pm \mathcal{E}_{s^{\prime}}=\mathcal{E}_{s^{\prime\prime}},\nonumber\\
&&E_3=0\Longrightarrow \sin k_1=0, M_1=\sqrt{(M_2\pm \mathcal{E}_3)^2+\lambda_2^2\sin k_2^2},\nonumber\\
&&E_4=0\Longrightarrow \sin k_1=0 , \sqrt{M_1^2+\mathcal{E}_2^2}=\mathcal{E}_3,\nonumber\\
&&E_5=0\Longrightarrow \mathcal{E}_2=0,\mathcal{E}_1=\mathcal{E}_3,\nonumber\\
&&E_6=0\Longrightarrow \sin k_1=0,\mathcal{E}_2=0,M_1=\mathcal{E}_3,\nonumber\\
&&E_7=0\Longrightarrow \mathcal{E}_3=\sqrt{\mathcal{E}_1^2+\mathcal{E}_2^2},\nonumber\\
&&E_8=0\Longrightarrow \sin k_1=0,\sin k_2=0,\sqrt{M_1^2+M_2^2 
    }=\mathcal{E}_3,\nonumber\\
&&E_9=0\Longrightarrow \sqrt{\mathcal{E}_1^2+\mathcal{E}_3^2\pm 2M_1\mathcal{E}_3}= \mathcal{E}_2,
\eeqn
where $s,s^{\prime},s^{\prime\prime}=1,2,3$.  Therefore, the bulk energy spectra of $\mathcal{H}_{1,6,8}$, $\mathcal{H}_{3,4,5}$, and $\mathcal{H}_{2,7,9}$ close at certain points, lines, and surfaces in the 3D Brillouin zone, respectively.

\end{appendix}

\bibliography{ref,reference1}

\end{document}